\documentclass[aps,prx,groupedaddress,superscriptaddress,twocolumn,floatfix]{revtex4-2}
\usepackage{amsmath}
\usepackage{graphicx}
\usepackage{float}
\usepackage{amssymb}
\usepackage[T1]{fontenc}
\usepackage{color,soul}
\usepackage[pdftex, pdftitle={PRX Energy test},pdflang={en-GB}]{hyperref} 
\usepackage{cleveref}
\usepackage[version=3]{mhchem} 

\usepackage[colorinlistoftodos]{todonotes}

\begin{document}

\title{Revisiting the storage capacity limit of graphite battery anodes: spontaneous lithium overintercalation at ambient pressure}

\author{Cristina Grosu}
\affiliation{Institute of Energy and Climate Research (IEK-9), Forschungszentrum J\"ulich, 52425 J\"ulich, Germany}
\affiliation{Chair for Theoretical Chemistry and Catalysis Research Center, Technical University of Munich, 85747 Garching, Germany}
\author{Chiara Panosetti} 
\email{panosetti@fhi.mpg.de} 
\affiliation{Chair for Theoretical Chemistry and Catalysis Research Center, Technical University of Munich, 85747 Garching, Germany}
\affiliation{Fritz-Haber-Institut der Max-Planck-Gesellschaft, 14195 Berlin, Germany}
\author{Steffen Merz} 
\affiliation{Institute of Energy and Climate Research (IEK-9), Forschungszentrum J\"ulich, 52425 J\"ulich, Germany}
\author{Peter Jakes}
\affiliation{Institute of Energy and Climate Research (IEK-9), Forschungszentrum J\"ulich, 52425 J\"ulich, Germany}
\author{Sebastian Matera} 
\affiliation{Fritz-Haber-Institut der Max-Planck-Gesellschaft, 14195 Berlin, Germany}
\affiliation{Institute for Mathematics, Freie Universit\"at Berlin, 14195 Berlin, Germany}
\author{R\"udiger-A. Eichel}
\affiliation{Institute of Energy and Climate Research (IEK-9), Forschungszentrum J\"ulich, 52425 J\"ulich, Germany}
\affiliation{RWTH Aachen University, Institute of Physical Chemistry, D-52074 Aachen, Germany}
\author{Josef Granwehr} 
\affiliation{Institute of Energy and Climate Research (IEK-9), Forschungszentrum J\"ulich, 52425 J\"ulich, Germany}
\affiliation{RWTH Aachen University, Institute of Technical and Macromolecular Chemistry, D‐52074 Aachen, Germany}
\author{Christoph Scheurer} 
\email{scheurer@fhi.mpg.de} 
\affiliation{Chair for Theoretical Chemistry and Catalysis Research Center, Technical University of Munich, 85747 Garching, Germany}
\affiliation{Fritz-Haber-Institut der Max-Planck-Gesellschaft, 14195 Berlin, Germany}
\date{\today}

\begin{abstract}

The market quest for fast-charging, safe, long-lasting and performant batteries drives the exploration of new energy storage materials, but also promotes fundamental investigations of materials already widely used. Presently, revamped interest in anode materials is observed -- primarily graphite electrodes for lithium-ion batteries. Here, we focus on the upper limit of lithium intercalation in the morphologically quasi-ideal highly oriented pyrolytic graphite (HOPG), with a \ce{LiC6} stoichiometry corresponding to 100\% state of charge (SOC). We prepared a sample by immersion in liquid lithium at ambient pressure and investigated it by static \ce{^7Li} nuclear magnetic resonance (NMR). We resolved unexpected signatures of superdense intercalation compounds, \ce{LiC_{$6-x$}}. These have been ruled out for decades, since the highest geometrically accessible composition, \ce{LiC2}, can only be prepared under high pressure. We thus challenge the widespread notion that any additional intercalation beyond \ce{LiC6} is not possible under ambient conditions. We monitored the sample upon calendaric aging and employed {\em ab initio} calculations to rationalise the NMR results. The computed relative stabilities of different superdense configurations reveal that non-negligible overintercalation does proceed spontaneously beyond the currently accepted capacity limit.
\begin{description}
\item

\end{description}
\end{abstract}
\keywords{Suggested keywords}
\maketitle
\tableofcontents

\section{Introduction}
\label{sec:intro}

For a reduction in greenhouse gas emission to tackle global warming, mass market penetration of electric vehicles (EVs) is a key element for a nearly \ce{CO2}-free transportation sector.\cite{United} Powerful, durable and safe lithium-ion batteries (LIBs) are crucial for consumer acceptance of electromobility on a larger scale. In particular, the fast-charging capability is regarded as a pivotal selling point. 
The necessity of fast-charging batteries brought some intrinsic limitations of the materials back into the spotlight, which historically did not matter in commonplace applications of LIBs, such as portable electronics. Primarily for the negative electrode, notable issues are still largely unaddressed to this day, including increasing the active site density, earlier detection of dendrite formation, or a quantitative description of mass and charge transport.\cite{Tarascon2001,Armand2008,Liu2019,Cai2020}
Extensive amount of work is ongoing to identify alternatives to the carbon-based anode materials. Even within the class of carbonaceous materials, different options were screened for usage as negative electrodes, from soft to hard carbon, carbon foam, carbon nanotubes, graphene sheets, artificial graphite, or mesocarbon microbeads graphite (MCMB).\cite{Asenbauer2020} Nonetheless, with its intrinsic capacity and wide availability, graphite is still the most employed anode material. Its working principle is based on the intercalation of lithium ions. Upon lithium intercalation during charging, graphite reaches its maximum reversible Li storage capacity at a lithium-to-carbon ratio of 1:6 (\ce{LiC6}). Theoretically this compound yields a capacity of 372~mAh/g, commonly defining 100\% state of charge (SOC).\cite{Dresselhaus2002,Asenbauer2020,Dahn1995} 
However, the highest geometrically accessible composition -- not considering lithium carbide (\ce{Li2C2}) but only the family of graphite intercalation compounds (GICs) -- is not \ce{LiC6}, but \ce{LiC2}, with a capacity three times higher. Nevertheless, the latter is metastable at ambient conditions and its non-electrochemical preparation was only reported under high pressure, with superdense decomposition products, \ce{LiC_{$6-x$}} with $x>0$, stable over an extended time period.\cite{Conard1994,Rabii2008}

\begin{figure}[t]
\includegraphics[width=.99\linewidth]{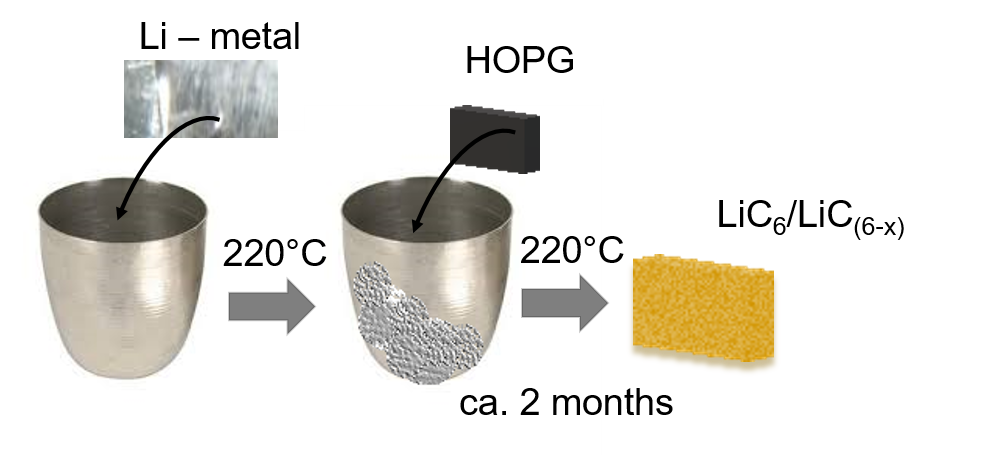}
\caption{\textbf{Schematic representation of the conducted synthesis.} First, metallic lithium was melted and kept at 220 $^{\circ}$C. Subsequently, a HOPG sample with dimensions of 10.0x(4.9)x2.0~mm was added. Tweezers were used to keep the HOPG in contact with molten lithium for a few minutes to start the intercalation. The system was then held for {\em ca.} ten weeks at constant conditions to ensure full intercalation into the HOPG.}
\label{Synthesis sketch}
\end{figure}

Despite extensive experimental efforts, in-depth understanding of {\em in operando} battery processes is still sparse.\cite{Pecher2016,Kayser2018,LorieLopez2018,Sacci2015} One major problem is that the interconnection of many relevant electrochemical processes renders the analysis of experimental data difficult and complicates the understanding of actual limits and causes of battery cell failure.\cite{Liu2019} Furthermore, theoretical investigations over large compositional Li/C ratios, and on time scales relevant for Li dynamics in graphite without unrealistic simplifications have only become possible recently. Therefore, concomitant theoretical and experimental research is scarce. 

Here, we use HOPG as a model system to investigate lithium ion intercalation, providing nuclear magnetic resonance (NMR) reference data from a system as well-defined as possible for further {\em in operando} studies of LIBs. To exclude any external influence on the intercalation process, we opted for an infiltration technique under ambient pressure.\cite{Conard1977,Guerard1975,Duan2020} Figure~\ref{Synthesis sketch} shows a sketch of the synthesis workflow, using lithium metal and HOPG as precursors. More details on the preparation are provided in the Methods section and in Supplementary Fig.~S1.a-b. Using this slow intercalation route, we obtained a sample that, repeatedly analysed in detail, showed spectroscopic evidence of superdense structures. We assessed the plausibility of such an assignment of spectral features by {\em ab initio} calculations. Eventually, we analysed the long-term evolution of the sample over several months, concluded by heating it up to 60~$^{\circ}$C.

\subsection{What do we know about superdense \ce{LiC2} and \ce{LiC_{6-x}}?}

\begin{figure}[t]
\includegraphics[width=.99\linewidth]{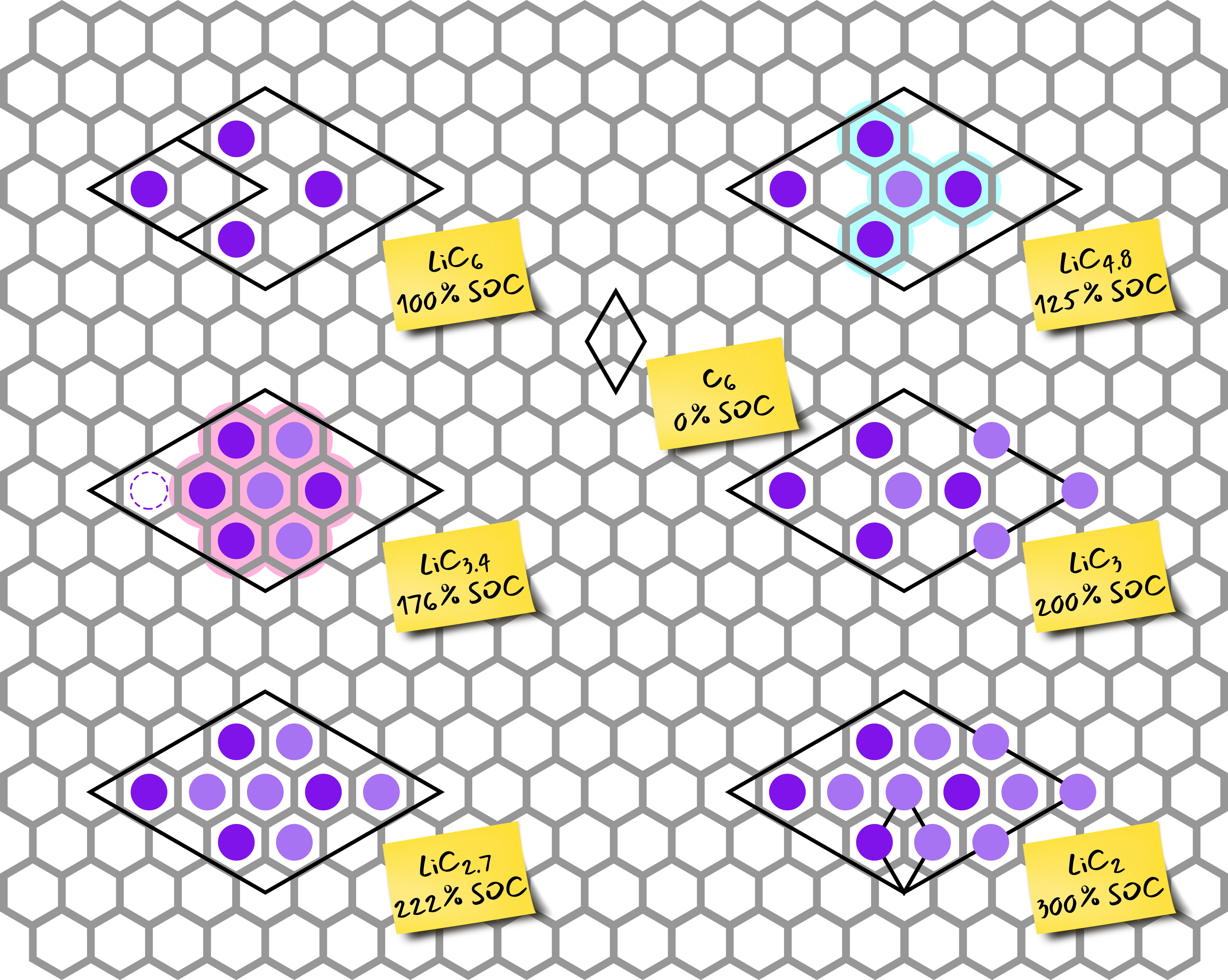}
\caption{\textbf{Schematic single-layer representation of some superdense \ce{LiC_{$6-x$}} high-symmetry structures and respective states of charge (SOC) relative to \ce{LiC6}.} The grey honeycomb represents the graphitic host lattice assumed to be AA stacked, and purple circles represent lithium atoms. Excess lithium with respect to \ce{LiC6} is represented in lighter purple. The dotted circle in \ce{LiC_{3.4}} marks how at least one Li atom must move from its original position in order to achieve the \ce{Li7} cluster motif. Black diamonds represent the smallest supercell commensurate to all the lithiated stoichiometries. For \ce{LiC6} and \ce{LiC2}, the (smaller) primitive cell is also shown. Highlighted in red and cyan are structural motifs that can give rise to high-ppm NMR signals.}
\label{structures}
\end{figure}

Superdense realisations of lithium GICs were reported at high pressure and temperature conditions since the 1980s.\cite{Schlke1989,Conard1994,Nalimova1995,Bindra1998,Bondarenko1998VibrationalSO,Schirmer1998,Guerard2004,Setton1994} Fig.~\ref{structures} schematically shows possible high-symmetry structures for some intermediate stoichiometries between \ce{LiC6} and \ce{LiC2}.

The occurrence of \ce{LiC2} was excluded {\em a priori} from any working battery for a long time. Since its non-electrochemical synthesis was always performed at high pressure and temperature, it is considered unlikely to be found in secondary batteries.\cite{Nalimova1995,Mordkovich1996,Rabii2008} 
Alternatively to high-pressure synthesis, ball-milling also allows \ce{LiC2} to be prepared using artificial graphite, MCMB, and carbon foam.\cite{Guerard2004,JANOT2005,Mordkovich1996} However, one may argue that ball-milling produces high pressure and temperature locally.\cite{Burmeister2018}
Once the pressure is released, the \ce{LiC2} composition becomes unstable and approaches \ce{LiC_{2.2-2.7,3.4}}.\cite{Nalimova1995,Nalimova1998} Nevertheless, \ce{LiC_{2.7}} and \ce{LiC_{3.4}} as decomposition products of \ce{LiC_2} were reported to be stable enough to allow measurements at ambient conditions.\cite{Bindra1998,Bondarenko1998VibrationalSO} Bindra {\em et al.} stabilised superdense GICs using boron doping, in an attempt to enable higher capacity electrode materials, yet they still used high-pressure synthesis.\cite{Bindra1998} 

For electrochemical intercalation, Conard {\em et al.} reported that the electric field actuates only as far as the gallery entrance, and therefore is not sufficiently attractive to drive the intercalation up to \ce{LiC2}.\cite{Conard2000}
Notwithstanding, superdense phases were observed for electrochemical systems using different carbon-based matrices, leading to the concept of overcharged LIBs.\cite{Hayes2003,Tossici2003,Azas2002}
In the latter works, lithium ions appeared to continue intercalating after \ce{LiC6} was formally reached, yet before being plated as lithium metal. Unfortunately, such overcharged anodes show irreversible capacity loss after the first deintercalation cycle.\cite{Tossici2003} More recently Paronyan {\em et al.} successfully used carbon foam as anode material to investigate overlithiation.\cite{Paronyan2017} 

A systematic investigation of the relative stability of overlithiated compounds, especially the intermediate stoichiometries between \ce{LiC6} and \ce{LiC2}, is currently missing, both from the experimental and the computational standpoint. On the experimental side, there is widespread consensus\cite{Avdeev1990, Reynier_2004, Reynier2007, Dre2013} that the free energy of intercalation of \ce{LiC6} falls in the range of -6 to -14~kJ/mol (-0.06 to -0.14~eV). Yet little is known about the free energies of overintercalation, except that the formation of \ce{LiC2} is assumed endergonic at ambient conditions. On the computational side, previous studies based on Density Functional Theory (DFT) only report total energy calculations of \ce{LiC6} and -- less often -- \ce{LiC2}.\cite{Imai2007, Rabii2008, Wang2014, Doh}. Two main limitations are recurring among these. First, dispersion interactions, which are crucial in intercalation chemistry, were not included in all the studies. Secondly, total energies only provide a ``virtual'' zero-temperature picture, while finite temperature and pressure require a description in terms of free energies. Despite these limitations, the reported total energies are compatible with the measured thermochemistry. However, no coherent body of literature exists that addresses the energetics of the entire \ce{LiC_{$6-x$}} range on the same footing.

\section{Materials and Methods} 
\label{sec:methods}

\subsection{Experimental}

\paragraph{Synthesis and sample preparation}
The \ce{LiC6} sample was prepared using an infiltration technique. Metallic lithium with 99.9~$\%$ purity (Sigma-Aldrich) was intercalated into highly oriented pyrolytic graphite HOPG (Goodfellow purchased by Sigma-Aldrich). The lithium metal was heated after the melting point up to 220~$^{\circ}$ C. The lithium's self-cleaning properties ensure a higher purity of the molten lithium metal. Afterwards the HOPG was added. In order to ensure complete lithium intercalation, the intercalation process has been allowed to take place for a period of over 2 months. This long infiltration time was necessary due to the dimension of the host material (pre-intercalation size = $10.0\times(4.9)\times2.0$ mm). \cite{Guerard1975,Conard1977,Duan2020}
The ageing process was performed in a closed container in a glove box under argon atmosphere. The sample was left to age for 5 months initially, followed by an additional period of 2 months. For the 5 months aged sample, a series of temperature dependency spectra of \ce{^7Li} static NMR were recorded. 

\paragraph{NMR measurements and data analysis}
All \ce{^7Li} NMR spectra were acquired using a Bruker BioSpin spectrometer Avance IIITM HD 600 XWB MHx at $B_{0}$ = 14.1 T (\ce{^7Li} Larmor frequency = 233.3 MHz) equipped with a Bruker DiffBB 5~mm BBO-H/F-Z Gradient diffusion probe-head. A single 90~$^{\circ}$ pulse excitation of 11~$\mu$s with recovery delays of 10~s on $^7$Li was employed.
The raw data were analysed using MATLAB. An exponential window function was employed before a fast Fourier transform (FFT). The spectra were zero- and first-order phase corrected as well as background corrected. 

\subsection{Computational}

\paragraph{DFT calculations and {\em ab initio} thermodynamics}
All the DFT calculations were performed using the plane-wave code {\tt{VASP}}\cite{Kresse1993} v.5.4.4, with the GGA-PBE functional\cite{Perdew1996} and the Projector Augmented-Wave (PAW) pseudopotentials.\cite{Kresse1999} Dispersion interactions were taken into account with the D3 method.\cite{Grimme2010} To ensure well converged total energies, a basis-set cutoff of 599~eV was chosen. All the calculated geometries were represented in appropriate periodic supercells and the Brillouin zone was sampled at a fixed k-point density of 0.1~\AA{}$^{-1}$. The vibrational densities of states were calculated using {\tt{phonopy}}\cite{phonopy} at the harmonic approximation level. The chemical potential of lithium was mapped to temperature in a 0-600~K range using Janaf thermochemical tables\cite{janaf} up to the fusion temperature (453.69~K) and employing a linear approximation for higher temperatures ({\em cf.} SI 2). AIMD simulations were performed by means of the DFTB\cite{elstner1998} code {\tt{dftb+}}\cite{Hourahine2020} v.19.1 using the parametrisation developed in our group\cite{Panosetti2021}. Trajectories of variable length were propagated with a time step of 1~fs in the canonical ensemble at 500, 750 and 1000~K, using a Nos\'e-Hoover thermostat\cite{Evans1985} with a coupling strength of 41~THz, corresponding to the highest vibrational mode of \ce{LiC6} as calculated with {\tt{phonopy}}.

\begin{figure*}[t]
\includegraphics[width=0.99\linewidth]{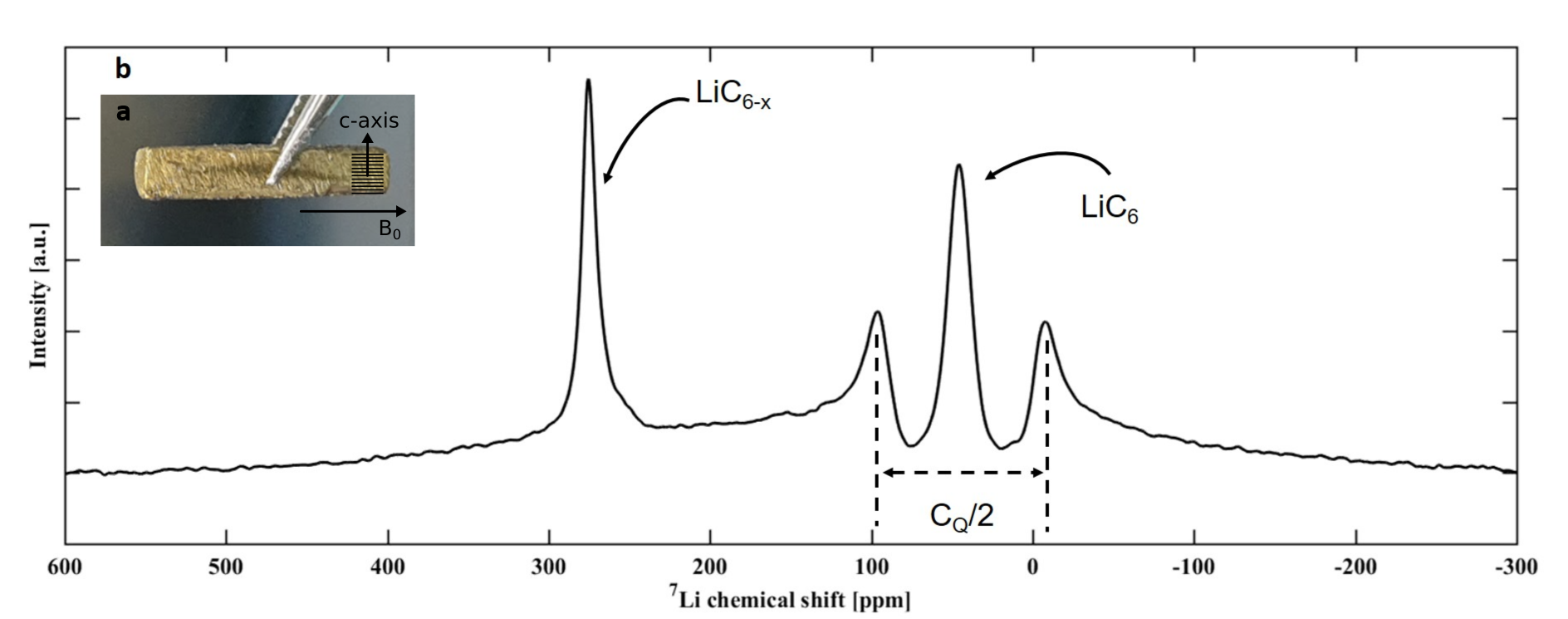}
\caption{\textbf{Static \ce{^7Li} NMR and initial sample in the NMR tube:} \textbf{a,} Shows the golden colour of lithium-intercalated HOPG, nominally corresponding to \ce{LiC6}.\cite{Guerard1975,Dresselhaus2002} Schematics in overlay show the orientation of the graphite layers, the crystallographic c-axis and the direction of $B_0$. 
\textbf{b,} Static \ce{^7Li} spectrum of lithium intercalated in HOPG. The isotropic chemical shift at 45 ppm is assigned to 
\ce{LiC6}. The quadrupolar satellites are compatible with a single-crystal pattern.\cite{Guerard1975,Conard2000}
The quadrupolar coupling constant $C_{Q}$ for \ce{LiC6} is 46 kHz. The peak at 274 ppm and the shoulder at 256 ppm show the presence of superdense \ce{LiC_{$6-x$}} compositions.\cite{Guerard2004}} 
\label{static nmr}
\end{figure*}

\section{Results and discussion}

\subsection{Static \ce{^7Li} nuclear magnetic resonance on lithium intercalated HOPG}

Static \ce{^7Li} NMR spectroscopy is a valuable tool to distinguish different degrees of lithiation in carbonaceous materials.\cite{Pecher2016} The \ce{^7Li} nucleus possesses spin 3/2, making observations of central transitions and quadrupolar satellite transitions possible. Satellite transitions allow conclusions on ordered structures with low Li-ion mobility,\cite{Guerard1975,Conard2000} whereas their absence may indicate a lack of perfectly ordered motifs and/or a motional averaging of the quadrupolar interaction.\cite{Conard2000} 

Fig.~\ref{static nmr}.a shows a photograph of the fully intercalated HOPG sample. The golden colour is characteristic of \ce{LiC6}. However, overlithiated compounds were also reported to appear golden.\cite{Bindra1998,Mordkovich1996}
Fig.~\ref{static nmr}.b represents the static \ce{^7Li} solid state NMR spectrum of the polished sample. All chemical shifts are reported against an external reference of a 1.0~M solution of LiCl in \ce{D2O}. We assign the isotropic chemical shift at 45~ppm with quadrupolar satellites to \ce{LiC6}, compatibly with previous works.\cite{Lauginie1992,Langer2013,LorieLopez2018} Additionally, an asymmetric signal with a sharp peak at 274 ppm and a shoulder at {\em ca.} 256 ppm is present. We also observe a broad spectral feature around 100-200 ppm, that is not background distortion ({\em cf.} SI~1). At first, the high-ppm signals may be attributed to plated lithium on the sample. The HOPG surface was polished using sand paper (Supplementary 1c-d), but residual metallic Li may persist. However, the intensity is too high for only trace amounts of lithium metal on the surface that are not detectable by visual inspection. Moreover, HOPG can be considered a defect-free single crystal, thus lacking internal pores that could accommodate pockets of metallic lithium.\cite{Winter98}

At this point, we must consider the possibility that these high-ppm signatures are generated by superdense phases instead. We are aware that superdense Li-GICs were never reported before without harsh pressure conditions. However, the uncertainty raised in our sample reflects the ambiguity of assignments present in literature. 
Aza\"is {\em et al.} assigned a 259~ppm \ce{^7Li} NMR resonance to lithium metal, despite using the exact ratio of Li/C to form \ce{LiC2} with a ball-milling synthesis. Interestingly, they found no Li metal signal in the X-ray spectrum, which is explained by intensive milling applied to the sample.\cite{Azas2002} 
Conversely, Conard {\em et al.} assigned a peak at 259 ppm to \ce{LiC2}.\cite{Conard1994} Since the spectrum was recorded after releasing the pressure, the true \ce{LiC2} chemical shift might be closer to the Li metal shift. The 259 ppm peak could then be attributed to \ce{LiC_{2.2-2.4}}.\cite{Conard1994,Nalimova1995,Nalimova1998}

As shown in Fig.~\ref{structures}, two local structural motifs are recurring in superdense compositions: a \ce{Li4} pattern in the shape of a three-pronged ``star'' (highlighted in cyan) and a denser \ce{Li7} pattern in the shape of a flat cluster, or ``flower'' (highlighted in red). The central Li atom is coordinated by three or six nearest Li neighbors, respectively. It is therefore expected to exhibit pseudo-metallic character. This was already suggested to explain the high-ppm signals associated to \ce{LiC2} decomposition products.\cite{Nalimova1995,Mordkovich1996,Conard2000}

Chang {\em et al.} showed that variable shifts at high-ppm can arise from different microstructures of lithium metal in electrochemical systems.\cite{Chang2015} Since no current or potential was applied during sample preparation, the formation of dendrites or mossy-type microstructures is not expected here. 
Trease {\em et al.} showed that lithium metal in non-spherical shapes is also sensitive to the direction of the magnetic field, with resonances shifting between about 245 and 270~ppm if a planar sample is placed perpendicular or parallel to $B_0$, respectively.\cite{Trease2012} This is known as orientation-dependent shift due to the bulk magnetic susceptibility effect. All our comparative measurements were performed using the same sample orientation, with the c-axis perpendicular to $B_{0}$ (Fig.~\ref{static nmr}). In a control experiment, we also cut part of the sample to measure at parallel orientation. The high-ppm signature only moved to 264 ppm (Supplementary Fig.~S2) within the limits set by the original 274 and 256~ppm signals. This indicates a susceptibility effect also for the quasi-metallic superdense species, as it has been observed for Li metal, before.

\subsection{Calendar ageing and post-ageing temperature dependent static \ce{^7Li} NMR of lithium intercalated HOPG}

\begin{figure}[t]
\includegraphics[width=0.99\linewidth]{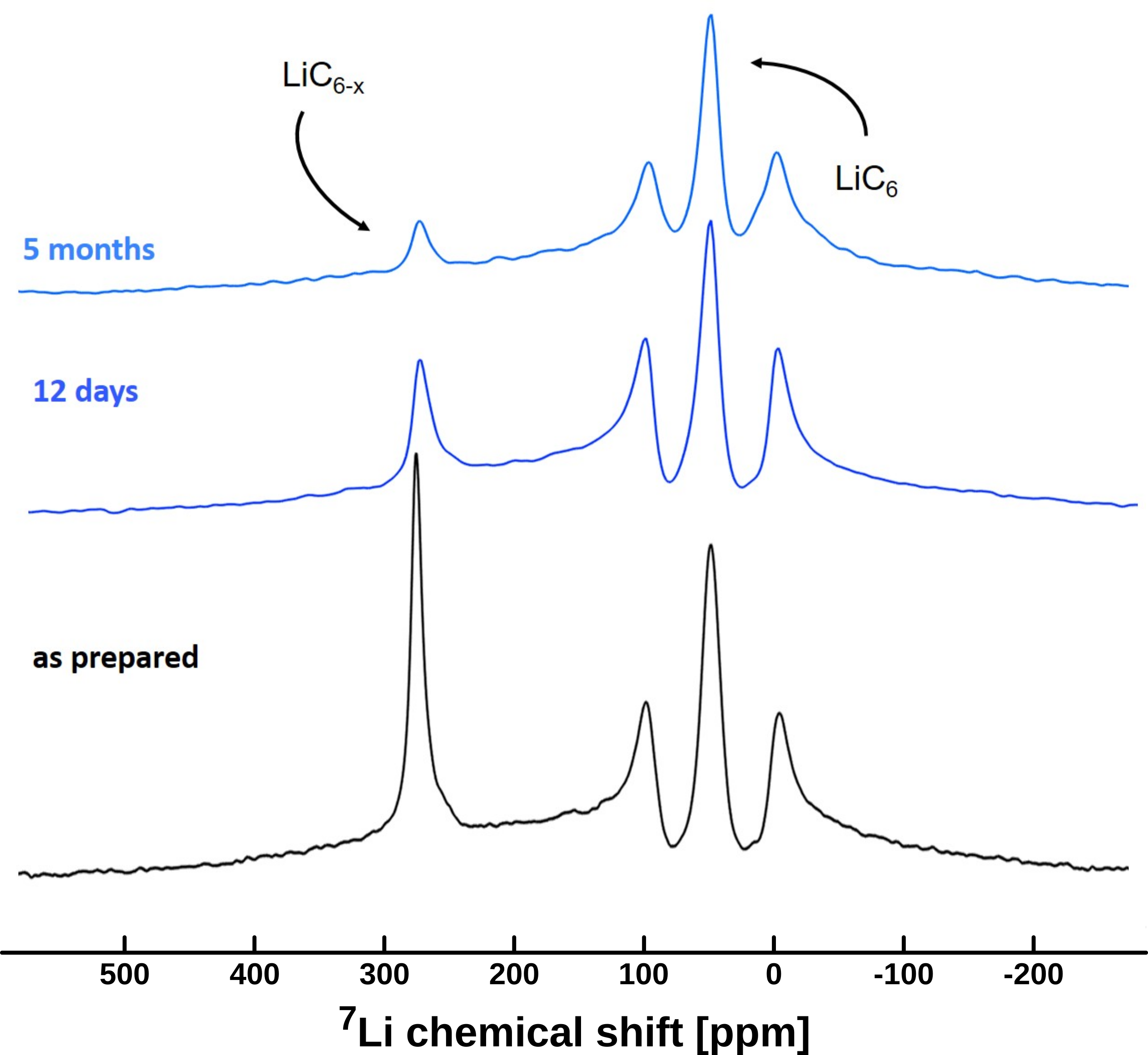}
\caption{\textbf{Evolution of the static \ce{^7Li} NMR spectrum during ageing of Li-intercalated HOPG.} The black curve shows the spectrum recorded after preparation. The dark blue curve shows the spectrum of the same sample aged for twelve days and the light blue curve after aging of five months. The signal appearing at 10-13\;ppm overlaps with the first quadrupole satellite peak of \ce{LiC6} and appears more pronounced after five months compared to twelve days.}
\label{NMR aged 3}
\end{figure} 

We aged the sample to investigate the long term \mbox{(meta-)stability} as well as changes in composition of \ce{LiC_{$6-x$}} over time. In addition, more invasive temperature dependent NMR experiments were performed after five months, followed by another two months of ageing.

\begin{figure}[t]
\includegraphics[width=0.99\linewidth]{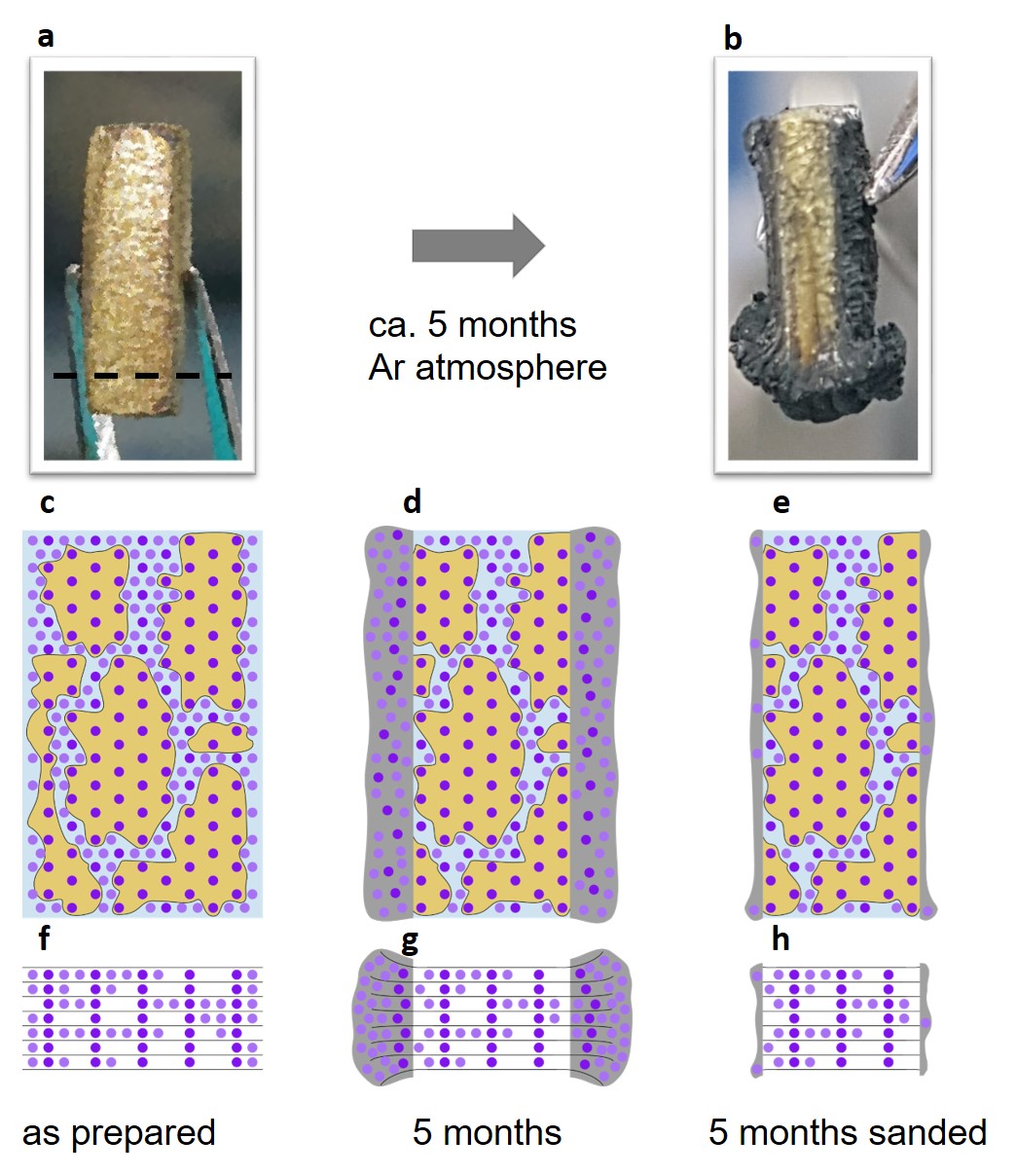}
\caption{\textbf{Representation of the sample: }\textbf{a,} Image of Li intercalated in HOPG after polishing. The golden colour indicates \ce{LiC6} or \ce{LiC_{$6-x$}} compounds. The dashed black line indicates the cut for complementary measurement ({\em cf.} Supplementary Fig.~S3). The golden face is shown after partially cleaning the sample. \textbf{b,} The sample after {\em ca.} five months of calendar aging in an inert atmosphere. The delamination shows the partial decomposition of the superdense compound. \textbf{c-h,} Schematic cuts through the overlithiated HOPG sample. Horizontal black lines in \textbf{f-h} represent graphene layers, and violet dots are Li nuclei. Golden shaded regions represent well-ordered \ce{LiC6} domains, giving rise to the 45 ppm NMR signal with quadrupolar satellites ({\em cf.} Fig.~\ref{static nmr}). The lower intensity of the satellites compared to the central transition indicates the presence of such ordered domains with limited dimensions, not spanning the whole HOPG sample. The blue interdomain region is overlithiated. Note that the dimensions are not drawn to scale -- the intensity of the satellites indicates \ce{LiC6} domains with a fairly large ratio of Li nuclei on the surface to Li in the volume of a domain of about 1:1. Upon aging, the surface of the lithiated HOPG crystal delaminates (grey shading), with disordered carbon forming that pulls Li from disordered regions, yet maintaining the \ce{LiC6} domains and the overlithiated interdomain region. Sanding only affects the surface, leaving the inner regions unchanged.}
\label{aged5months}
\end{figure}

\begin{figure}[t]
\includegraphics[width=0.99\linewidth]{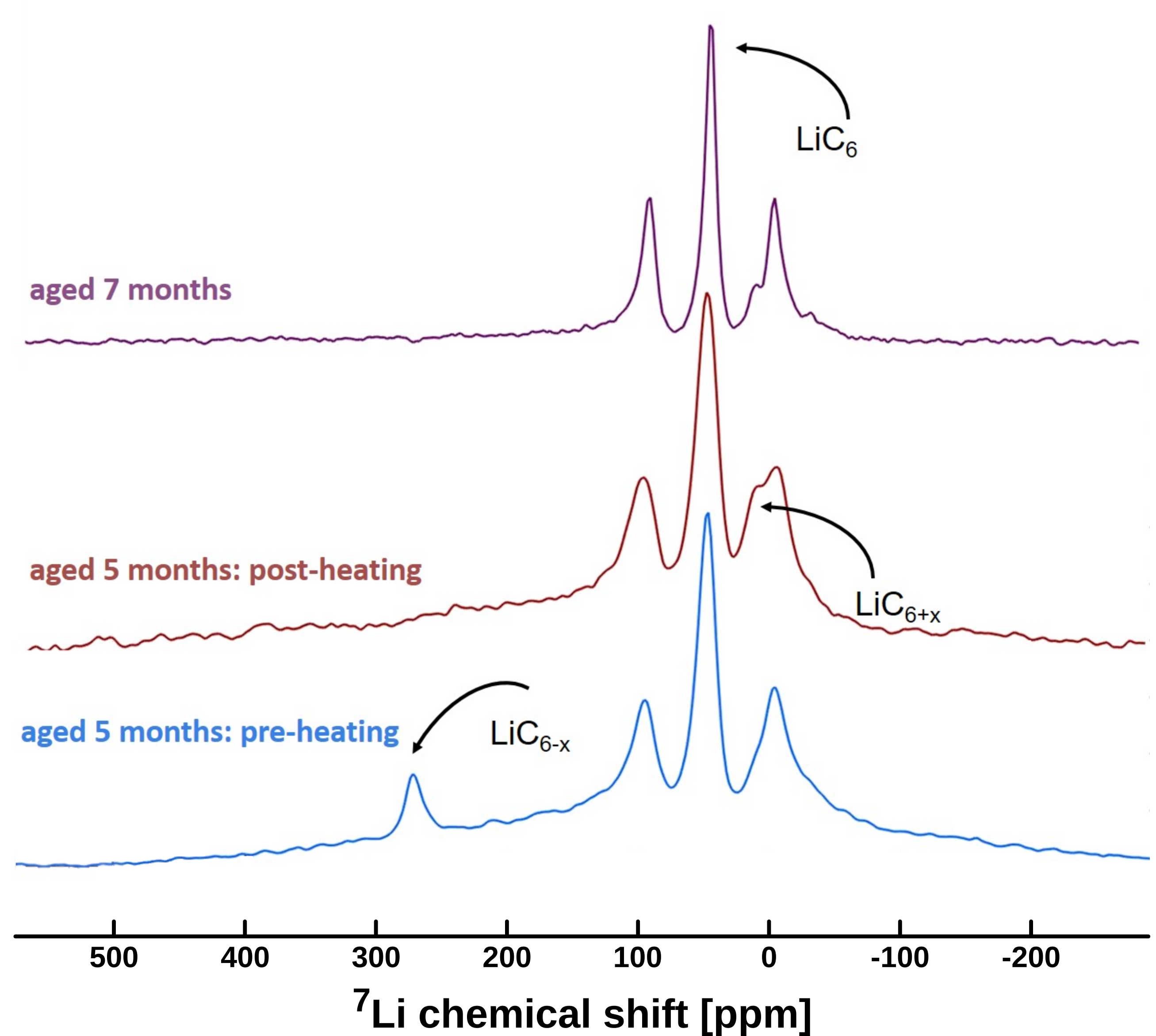}
\caption{\textbf{Static {\ce{^7}Li} NMR spectra taken after the decomposition process and pre- and post-heating to accelerate the equilibration of the residual superdense phases.} All curves are shown for 25\;$^\circ$C, while the sample was treated and measured from $-10\;^\circ$C to $60\;^\circ$C ({\em cf.} Supplementary Fig.~S3). The violet curve shows the sample aged for two additional months. The peak at 10-13\;ppm remains stable.}
\label{NMR aged}
\end{figure}

We recorded static \ce{^7Li} NMR spectra after twelve days (dark blue curve in Fig.~\ref{NMR aged 3}) and after five months (light blue curve Fig.~\ref{NMR aged 3}). The sample was thoroughly cleaned before each measurement, exposing golden shiny faces on each side of the HOPG crystal. We observe a decrease in the high-ppm peak intensities, suggesting a partial degradation of the corresponding structures. The degradation process appeared incomplete, with a residual broad signal that appears as the overlap of both shifts present in the fresh sample spectrum. The stability of the structures associated with these signals over several months is compatible with Nalimova {\em et al.}\cite{Nalimova1995} The 274 ppm signal seems to decrease the most during this timeframe but cannot be fully disentangled from the 256~ppm (light blue {\em vs.} black curve Fig.~\ref{NMR aged 3})
A new Li environment with a chemical shift of 10--13~ppm forms concomitantly. Signatures in this region are commonly associated to lower SOC and generically identified as \ce{LiC_{$6+x$}},\cite{LorieLopez2018} or, in other works, attributed to Li--Li dimers.\cite{Sato1994}

After five months the sample decomposed visibly as shown in Fig.~\ref{aged5months}. The observed drastic delamination is only compatible with the expulsion of lithium from inside the sample, thus it would not have occurred if the high-ppm signal was caused by surface metal only. The opening of the graphite sheets also indicates pressure release from within the bulk material.\cite{Nalimova1995,Mordkovich1996,Nalimova1998} This is a further indicator towards the degradation of a superdense structure. Fig.~\ref{aged5months} also shows a schematic model of the possible microscopic configuration of sample (Panels a, f), with ordered \ce{LiC6} domains and overlithiated interdomain regions. Panels d, g show the schematic delamination after five months. Panels e, h represent the sample after cleaning.

In the temperature dependent measurements (Fig.~\ref{NMR aged} and Supplementary Fig. S3), the residual high-ppm peak disappears after mild heating to 310--330~K, while the 10--13 ppm signal becomes more evident ({\em cf.} Supplementary Fig.~S3). The broad spectral feature between about 100--200~ppm, which temporarily vanishes at elevated temperature during the heating cycle ({\em cf.} Supplementary Fig.~S3 at 40 and 60°C), reappears post-heating and only vanishes permanently after two more months of ageing. 

\begin{table*}[t]
\centering
\caption{\label{tab:aitd} Formation energies of selected \ce{LiC_{$6-x$}} intercalation compounds. $\Delta E$ is calculated from DFT total energies ({\em cf.} Methods section), not including the zero point energies nor any finite temperature contribution. The values of $\Delta G$ at different temperatures and ambient pressure are calculated using the {\em ab initio} thermodynamics approach. Our experimental conditions are T = 500 K and P = 1 atm. In bold, non-negligible populations at 300 and 500~K.}
\begin{ruledtabular}
\begin{tabular}{ccccccc}
 & \ce{LiC6} & \ce{LiC_{4.8}} & \ce{LiC_{3.4}} & $\ce{LiC_{3}}$ & \ce{LiC_{2.7}} & \ce{LiC2} \\
\hline
$\Delta E^{\mathrm{interc}}$ (DFT-D3) / eV & -0.10 & -0.01 & 0.14 & 0.27 & 0.46 & 1.08 \\
$\Delta G^{\mathrm{interc}}$ (300 K) / eV & -0.12 & -0.03 & 0.09 & 0.24 & 0.40 & 1.00 \\
$\Delta G^{\mathrm{interc}}$ (500 K) / eV\ & -0.09 & 0.02 & 0.12 & 0.28 & 0.44 & 1.09 \\
\hline
$ N(\ce{LiC_{6-x}})/N(\ce{LiC6})$ (300 K) / eV & --- & \bf{0.03} & 0.00 & 0.00 & 0.00 & 0.00 \\
$ N(\ce{LiC_{6-x}})/N(\ce{LiC6})$ (500 K) / eV & --- & \bf{0.08} & \bf{0.01} & 0.00 & 0.00 & 0.00
\end{tabular}
\end{ruledtabular}
\end{table*}

\subsection{{\em Ab initio} thermodynamics and dynamics of superdense GICs}

A basic modelling approach is to initially consider periodic high-symmetry structures as shown in Fig.~\ref{structures}, and calculate the formation free energies of these extended ``pure'' phases. To this end, we adopted an {\em ab initio} thermodynamics (AITD) approach.\cite{Rogal2006} A detailed derivation of the formalism adapted to our system is provided in SI 2. Within this framework we calculated the free energy of intercalation $\Delta G^{\mathrm{interc}}$ for the stoichiometries above at 300 and 500~K and ambient pressure. Based on these, we estimate relative populations $N(\ce{LiC_{6-x}})/N(\ce{LiC6})$ at equilibrium as Boltzmann ratios with respect to \ce{LiC6}. The results are reported in Table~\ref{tab:aitd}. The effect of configurational entropy was neglected, which would further favour overlithiated compounds (except \ce{LiC2}; {\em cf.} SI 2). As such, the relative populations are to be considered a lower estimate.

With \ce{LiC2} at +1.09 eV (105.17 kJ/mol) at 500~K, we exclude its presence in the sample. However, \ce{LiC_{3.4}} and \ce{LiC_{4.8}} have only mildly positive $\Delta G^{\mathrm{interc}}$ at 500~K, which becomes even negative at room temperature for \ce{LiC_{4.8}}. Correspondingly, their relative populations are non-vanishing.
Regardless of the precise assignment of the high-ppm signal ({\em vide infra}), we stress at this point that the energetics above, albeit simplified, confirm that a certain amount of excess lithium does indeed enter spontaneously. In other words, the common conception that \ce{LiC6} corresponds to 100~\% SOC is not entirely accurate thermodynamically. The exact upper limit of overlithiation accessible beyond \ce{LiC6} and its dependence on external conditions can only be determined by means of computationally expensive statistical sampling, which goes beyond the scope of this work.

The observed high-ppm signal may indeed arise from \ce{Li7} ``flowers'' and/or \ce{Li4} ``stars'' present not only in sizeable domains of exact \ce{LiC_{3.4}} and \ce{LiC_{4.8}} compositions, but also diluted in a \ce{LiC6} environment. Considering that, starting from a \ce{LiC6} environment, every additional lithium will form at least a \ce{Li4} ``star'', this can happen at any \ce{LiC_{$6-x$}} stoichiometry. 
Additionally, a ``flower'' may form if three neighbouring Li atoms from the immediate surroundings aggregate around the centre of a ``star''. We estimate the cost of such aggregation as {\em ca.} 0.11~eV, thus also thermally accessible. Subsequently, ``star'' and ``flower'' motifs are in equilibrium with each other, thus both can contribute to the high-ppm signals.

The relative populations at 300~K ({\em cf.} Table~\ref{tab:aitd}) confirm the metastability of superdense patterns at room temperature. However, the relative concentrations of these patterns in the fresh sample must be closer to those at 500~K than those at 300~K, as the calendaric aging showed a slow equilibration towards degradation.
Assuming that the decomposition is diffusion-controlled, we estimate the effective diffusion barrier and thus the relative degradation rate at 300 {\em vs.} 500 K. We performed {\em ab initio} molecular dynamics (AIMD) simulations, based on Density Functional Tight Binding (DFTB)\cite{elstner1998}, to evaluate Li mobility in a slightly oversaturated \ce{LiC_{$6-x$}} supercell with two \ce{Li7} motifs in an \ce{LiC6}-like environment. The resulting diffusion coefficients (Supplementary Fig.~S4) show Arrhenius behaviour with an effective barrier of 0.35~eV, which slows down the delithiation about 225 times at room temperature compared to 500 K ({\em cf.} SI 3). 

We note in passing that the AIMD trajectories exhibited frequent occurrences of directly connected Li dimers and trimers in an isosceles triangular configuration as transient byproducts of the decomposition of the \ce{Li7} clusters (Supplementary Fig.~S5). This nicely ties in with the appearance of the low-ppm shoulder upon ageing, if this were to be attributed to Li--Li dimers rather than low-SOC patterns. 

\subsection{Discussion}
Without explicitly simulating chemical shifts, we cannot unambiguously assign each resonance. We excluded that the high ppm features are generated by lithium metal. On energetic grounds, we also exclude full \ce{LiC2}. Thus, we infer the presence of some form of intermediately overlithiated phase containing \ce{Li7} (``flowers'') and/or \ce{Li4} (``star'') motifs, and associate the high-ppm shift to pseudo-metallic character of the central atoms. While one may argue whether the \ce{Li4} stars are ``dense'' enough to produce high-ppm shifts, their formation is more energetically accessible than that of \ce{Li7} flowers, therefore a spectroscopic signature is to be expected. 
In the light of this, we put forward the following scenario, qualitatively combining experimental observations and simulations. The asymmetric high-ppm signal corresponds to sizeable domains of at least \ce{LiC_{4.8}} stoichiometry, possibly mixed with \ce{LiC_{3.4}}, and domains of variable \ce{LiC_{$x>3.4$}} stoichiometries with \ce{Li7} and/or \ce{Li4} motifs diluted in \ce{LiC6}-like surroundings. The signal is split into a sharp 274 ppm feature and a broad 256 feature due to susceptibility effects analogously to metallic lithium, but of smaller magnitude. As such, the more intense 274 ppm peak would correspond to sample faces with normal vectors perpendicular to $B_0$, while the 256 ppm corresponds to face normals parallel to $B_0$. It is reasonable to assume that overlithiation mainly occurs in the proximity of the HOPG surface, while the inner bulk is predominantly \ce{LiC6}. Then, the relative magnitude of the two peaks reflects the dimensions of the sample (longer surface along $B_0$). Moreover, as the core of the sample may be shielded to a fair degree, surface species may be weighted more strongly and therefore show a higher relative amplitude with respect to the \ce{LiC6} bulk than relative energetics would suggest.\cite{nla.cat-vn1860681}
Considering the inherent metastability of \ce{Li7} clusters, an additional significant population of imperfect \ce{Li_$n$} clusters (with $3 \geq n > 7$) can be expected (``broken flowers''). The lithium atoms belonging to these clusters are undercoordinated with respect to the \ce{Li7} central atom but still occupying adjacent \ce{C6} rings, therefore they can be expected to produce a signal at a higher shift than the ``free'' lithium atoms in \ce{LiC6}. \ce{Li3} motifs can analogously appear as ``broken stars''. Additionally, both the crown atoms of the \ce{Li7} and the prongs of the \ce{Li4} motifs have lower coordination than the respective central atoms. As such, there are many possible realisations of microstructures with a wide range of coordinations, thus we may attribute the broad spectral feature at 100--200~ppm to a superposition of resonances corresponding to all the above. 
Both the high-ppm peaks and the broad 100--200~ppm spectral feature are correlated with the increase of the 10--13 ppm feature. This is a strong indication that the degradation of superdense structures directly corresponds to the appearance of a new Li environment. In line with previous assignments of the low-ppm feature in literature, this can correspond to either Li--Li dimers occurring as the smallest possible decomposition product before isolated Li, or to the formation of locally Li-depleted \ce{LiC_{$6+x$}} regions following the ejection of Li at the surface. Of note, these two patterns can coexist as a result of decomposition (cf. Supplementary Fig.~S5).

\section{Conclusions}

While preparing a reference sample of fully intercalated \ce{LiC6} for \ce{^7Li} NMR spectra, we observed unexpected high-ppm resonances. Confidently ruling out that the observed signatures arise from residual metallic lithium, we attribute these to superdense \ce{LiC_{$6-x$}} compounds formed under ambient pressure. We investigate the evolution of the signal under calendaric aging and rationalise our observations with {\em ab initio} simulations. We infer that the signal arises from sizeable domains containing \ce{Li7} (``flowers'') and/or \ce{Li4} (``stars'') motifs in sufficient amounts and we estimate the long-term \mbox{(meta-)stability}. {\em Ab initio} thermodynamics confirms that a non-negligible excess of lithium enters spontaneously, which, to the best of our knowledge, had never been considered before. 
These findings challenge the currently accepted hypothesis that, since \ce{LiC2} can only be prepared under high pressure\cite{Nalimova1995,Mordkovich1996,Conard2000}, any additonal intercalation beyond \ce{LiC6} is implausible. In hindsight, the simple consideration that the range of stoichiometries between \ce{LiC6} and \ce{LiC2} spans 200\% states of charge beyond 100\% should suggest prudence in such an assumption. Yet, it was never rigorously verified. In our view, multiple previous works on electrochemical cells contain indications compatible with at least a sparkle of doubt.\cite{Hayes2003,Tossici2003,Paronyan2017} To be fair, the question of assessing the true capacity of ordered graphitic hosts was explicitly addressed for bilayer or multilayer graphene.\cite{Khne2018,Ji2019} Inexplicably however, the evidence of overlithiation in the latter did not reopen the question of analogous occurrence in extended graphite -- which is the material actually used in working batteries. For decades, superdense graphite intercalation compounds have been considered only accessible as decomposition products of \ce{LiC2} under high-pressure synthesis (``from above'').\cite{Nalimova1995,Mordkovich1996,Conard2000} Here we confirm that superdense compositions are also accessible directly as overintercalation products of \ce{LiC6} at ambient pressure (``from below''). If this is possible under the synthesis conditions employed here, it is reasonable to expect that overlithiation is further favoured under an applied potential. Particularly in fast charging conditions, lithium plating is also increasingly favoured.\cite{Wandt18} Hence, the most intriguing aspect is the interplay of partially reversible plating, overlithiation and reintercalation. On this account, we call for a reconsideration of the role of overlithiation, so far excluded from the picture with graphite as a host. Taking overlithiation into consideration may also shed light onto other hitherto unexplained phenomena, such as the apparent ``disappearance'' of some amount of available lithium between cycles, the latter commonly attributed solely to the formation of solid electrolyte interfaces (SEI) and ``dead'' lithium.\cite{Sole2014} Finally, With regards to the use of NMR for the detection of plating and dendrite formation, our results also suggest that caution is indicated with assigning high-ppm signals non-specifically to the emergence of metallic lithium deposits. 



\section*{Acknowledgements}
All the authors thank Philipp M. Schleker and Svitlana Taranenko for valuable contribution to experimental aspects of this project and for the valuable discussion on the NMR and Matthias Kick and Simon Anni\'es for fruitful discussions on the theory and data presentation. This work was funded by the German Federal Ministry of Education and Research (BMBF) as part of the research cluster ``AQua'' within the project InOPlaBat (grant number 03XP0352) and by the research initiative Jülich Aachen Research Alliance (JARA), section JARA-Energy, within the project MF 001-17 (project ID: G:(DE-82)ZUK2-SF-JARA-ENERGY MF 001-17). The authors gratefully acknowledge the computational and data resources provided by the Leibniz Supercomputing Centre (LRZ).



\bibliography{bibliography}

\end{document}


\title{Supplementary Information: Revisiting the storage capacity limit of graphite battery anodes: spontaneous lithium overintercalation at ambient pressure}

\author{Cristina Grosu}
\affiliation{Institute of Energy an Climate Research (IEK-9), Forschungszentrum J\"ulich, 52425 J\"ulich, Germany}
\affiliation{Chair for Theoretical Chemistry and Catalysis Research Center, Technical University of Munich, 85747 Garching, Germany}
\author{Chiara Panosetti} 
\email{panosetti@fhi.mpg.de} 
\affiliation{Chair for Theoretical Chemistry and Catalysis Research Center, Technical University of Munich, 85747 Garching, Germany}
\affiliation{Fritz-Haber-Institut der Max-Planck-Gesellschaft, 14195 Berlin, Germany}
\author{Steffen Merz} 
\affiliation{Institute of Energy an Climate Research (IEK-9), Forschungszentrum J\"ulich, 52425 J\"ulich, Germany}
\author{Peter Jakes}
\affiliation{Institute of Energy an Climate Research (IEK-9), Forschungszentrum J\"ulich, 52425 J\"ulich, Germany}
\author{Sebastian Matera} 
\affiliation{Institute for Mathematics, Freie Universit\"at Berlin, 14195 Berlin, Germany}
\affiliation{Fritz-Haber-Institut der Max-Planck-Gesellschaft, 14195 Berlin, Germany}
\author{R\"udiger-A. Eichel}
\affiliation{Institute of Energy an Climate Research (IEK-9), Forschungszentrum J\"ulich, 52425 J\"ulich, Germany}
\affiliation{RWTH Aachen University, Institute of Physical Chemistry, D-52074 Aachen, Germany}
\author{Josef Granwehr} 
\affiliation{Institute of Energy an Climate Research (IEK-9), Forschungszentrum J\"ulich, 52425 J\"ulich, Germany}
\affiliation{RWTH Aachen University, Institute of Technical and Macromolecular Chemistry, D‐52074 Aachen, Germany}
\author{Christoph Scheurer} 
\email{scheurer@fhi.mpg.de} 
\affiliation{Chair for Theoretical Chemistry and Catalysis Research Center, Technical University of Munich, 85747 Garching, Germany}
\affiliation{Fritz-Haber-Institut der Max-Planck-Gesellschaft, 14195 Berlin, Germany}
\date{\today}
\keywords{}
\maketitle

\section{SI 1. Experimental Section}

\subsection{Sample preparation}
\begin{figure}[h]
\includegraphics[width=.95\linewidth]{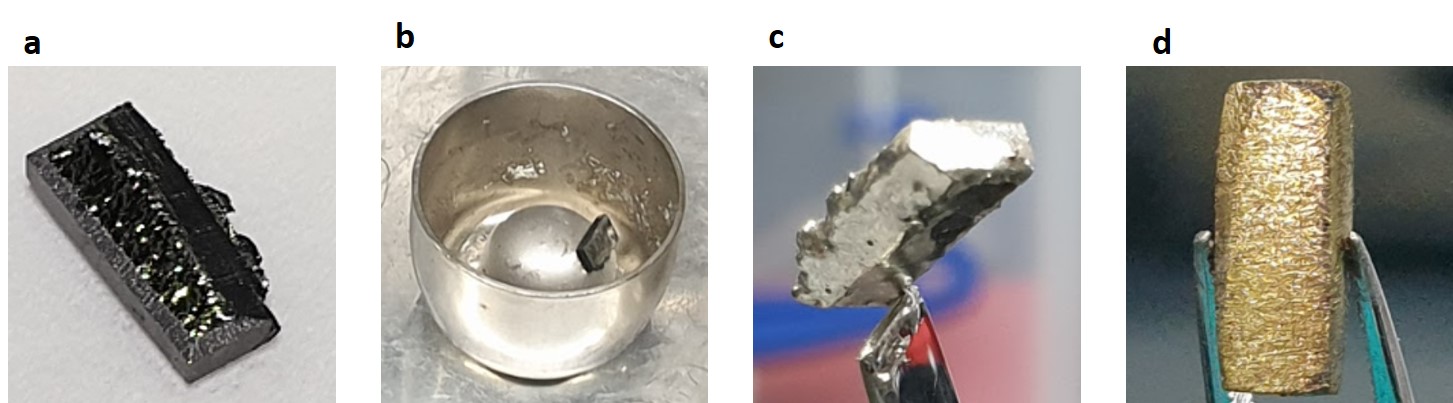}
\caption{\textbf{Summary of the steps to produce the lithium--graphite intercalation compound.} \textbf{a,} shows the raw sample of HOPG, cut mechanically from a 10.0x10.0x2.0~mm$^3$ piece. \textbf{b,} represents the molten lithium in a nickel crucible and the subsequently inserted HOPG, whereas \textbf{c,} is the fully intercalated sample after {\em ca.} two months, plated by lithium metal, as it was removed from the molten lithium at the end of the synthesis. \textbf{d,} shows the final stage post polishing. The golden colour indicates full lithiation. 
}
\label{materialprep}
\end{figure}

The \ce{LiC6} sample was prepared using an infiltration technique. Metallic lithium with 99.9~\% purity (Sigma Aldrich) was intercalated in highly oriented pyrolytic graphite (HOPG) from Goodfellow. The lithium metal was heated above the melting point until 220 $^{\circ}$C was reached. Note that the self-cleaning property of lithium ensures an even higher purity of the liquid lithium metal bulk.\cite{Duan2020} Afterwards the HOPG was added. As one can see in Fig.~\ref{materialprep}.a, the sample was cut, in order to expose some edges and to facilitate the starting of the intercalation by maximizing lithium wetting on graphite, although recent studies show that graphite is lithiophilic.\cite{Duan2020}
In order to ensure complete lithium intercalation, the high-temperature intercalation process has been allowed to take place for a period of over 2 months in an inert atmosphere. This long infiltration time was necessary due to the dimension of the host material pre-intercalation size of 10.0x(4.9)x2.0~mm$^3$.\cite{Guerard1975,Conard1977,Duan2020} 
The dimensions were determined by the requirement to fit the final sample into an NMR tube of 5~mm diameter. The bracket notation is due to the cut of the original 10.0x10.0x2.0~mm$^3$ piece. The final sample was, by applying a cleaning procedure, adapted to fit within the NMR tube, so reduced even more in dimension. 
This was done, before recording the nuclear magnetic resonance (NMR) spectra, by polishing the fully intercalated HOPG mechanically using sand paper, in order to avoid further chemical contamination. 
The full synthesis workflow is sketched in Fig.~\ref{materialprep}.

\subsection{\ce{^7Li} nuclear magnetic resonance}
\begin{figure}
\includegraphics[width=.95\linewidth]{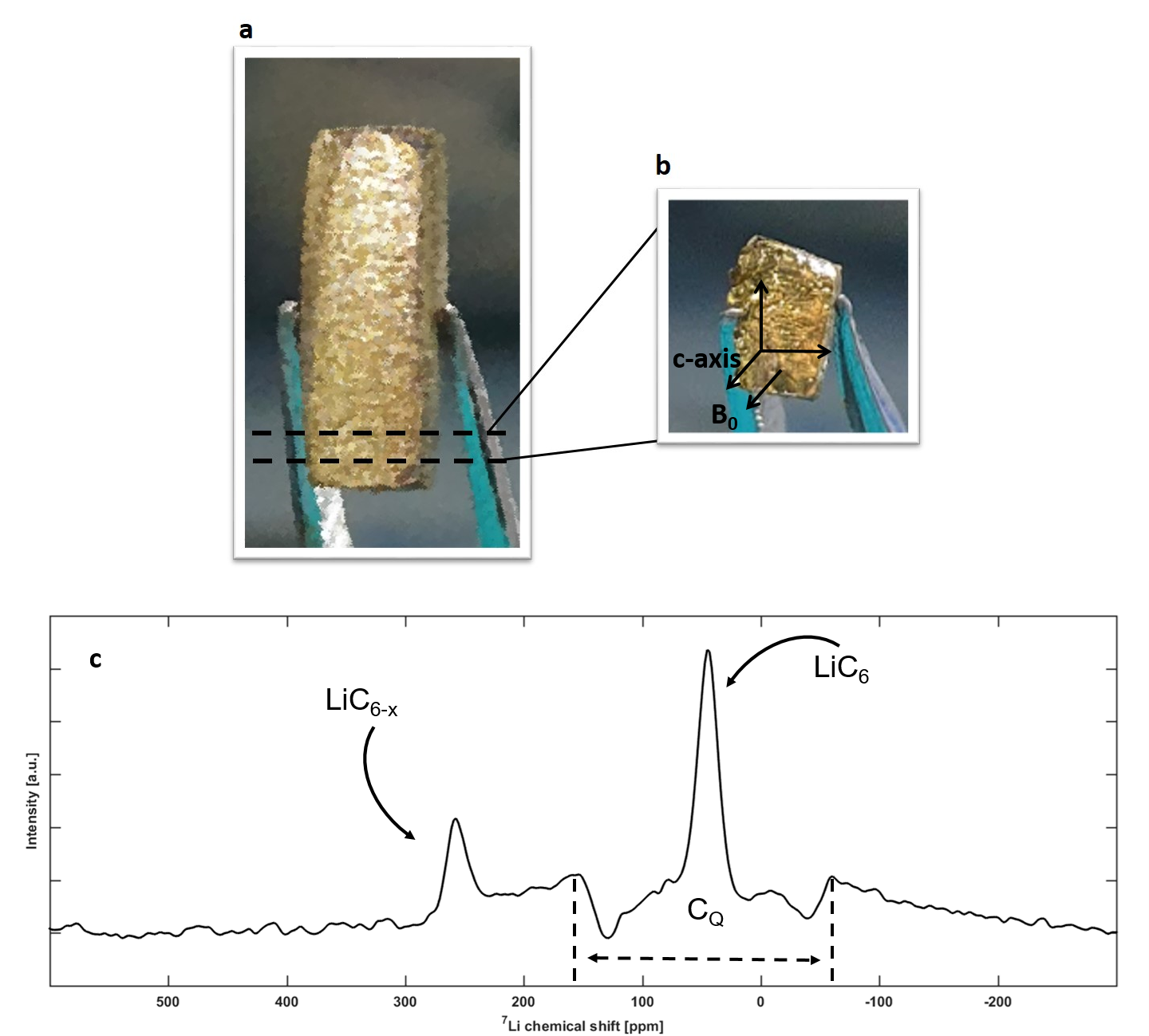}
\caption{\textbf{Photograph of the fully intercalated sample and the cut out fraction for the in-plane \ce{^7Li} NMR measurements.} \textbf{a,} showing the whole sample after mechanical cleaning; the black dashed lines indicate the location of the cut to produce a sliced sample that would fit into an NMR tube with its c-axis rotated by $90^\circ$ with respect to the external magnetic field $B_0$. \textbf{b,} displays the sliced part from the main sample for measurement at the parallel orientation of the c-axis towards the $B_0$ field. The a and b directions are not labeled. \textbf{c,} \textsuperscript{7}Li NMR spectrum of the sample with its c-axis oriented parallel to the $B_0$ field direction. The signature at 45~ppm corresponds to the \ce{LiC6} phase, while the peak at 264~ppm, excluding a lithium metal signal (see main text), corresponds to the \ce{LiC_{$6-x$}} phase.}
\label{rotated}
\end{figure}

Susceptibility effects are known to be present when measuring the $^7$Li NMR spectra of metallic or pseudo-metallic samples. In particular, lithium metal can exhibit shift variations of {\em ca.} 10--30~ppm, depending on the sample geometry and the angle of the c-axis of the sample with respect to the $B_0$ field direction. The maximum difference will be found comparing sample orientations with surface normal vectors aligned parallel and perpendicular to $B_0$.\cite{Trease2012,Bhattacharyya2010}
To investigate the influence of the susceptibility, our sample was cut as shown in Fig.~\ref{rotated}.a-b. Fig.~\ref{rotated}.c shows the \ce{^7Li} NMR spectrum, with the graphite c-axis oriented parallel to the $B_0$ field, while Fig. 3 in the main text (as well as Fig.~S3) shows the spectrum taken at perpendicular c-axis orientation towards $B_0$. The signal-to-noise ratio (S/N) is weaker due to the smaller size of the sample. The high-ppm chemical shift is moving from 274~ppm for perpendicular orientation to 264~ppm for parallel orientation. The effect is similar to that observed for a Li metal sheet, but of different magnitude (10 {\em vs.} 30~ppm, respectively).\cite{Trease2012} However, given the different shapes of the two samples (Fig.~\ref{rotated}) and the anisotropic nature of susceptibility in graphite, no quantitative conclusion can be drawn at this point. Since our main conclusions are reached by comparing a sample with invariant shape at identical orientation, this aspect is not characterized in more detail here.

By comparing the resonance frequencies of the satellite transitions of the resonance at 45~ppm for our single-crystal sample placed with its crystallographic c-axis
in perpendicular and parallel directions with respect to the magnetic field $B_0$, the quadrupolar coupling constant can be extracted. The results are consistent with the values reported by Roth {\em et al.} \cite{Roth1981}, with Fig.~\ref{rotated}.c showing the full quadrupolar constant $C_Q$ at {\em ca.} 47 kHz and Fig.~3.b in the main text shows $C_{Q/2}$ with {\em ca.} 23 kHz.

\begin{figure}
\includegraphics[width=.95\linewidth]{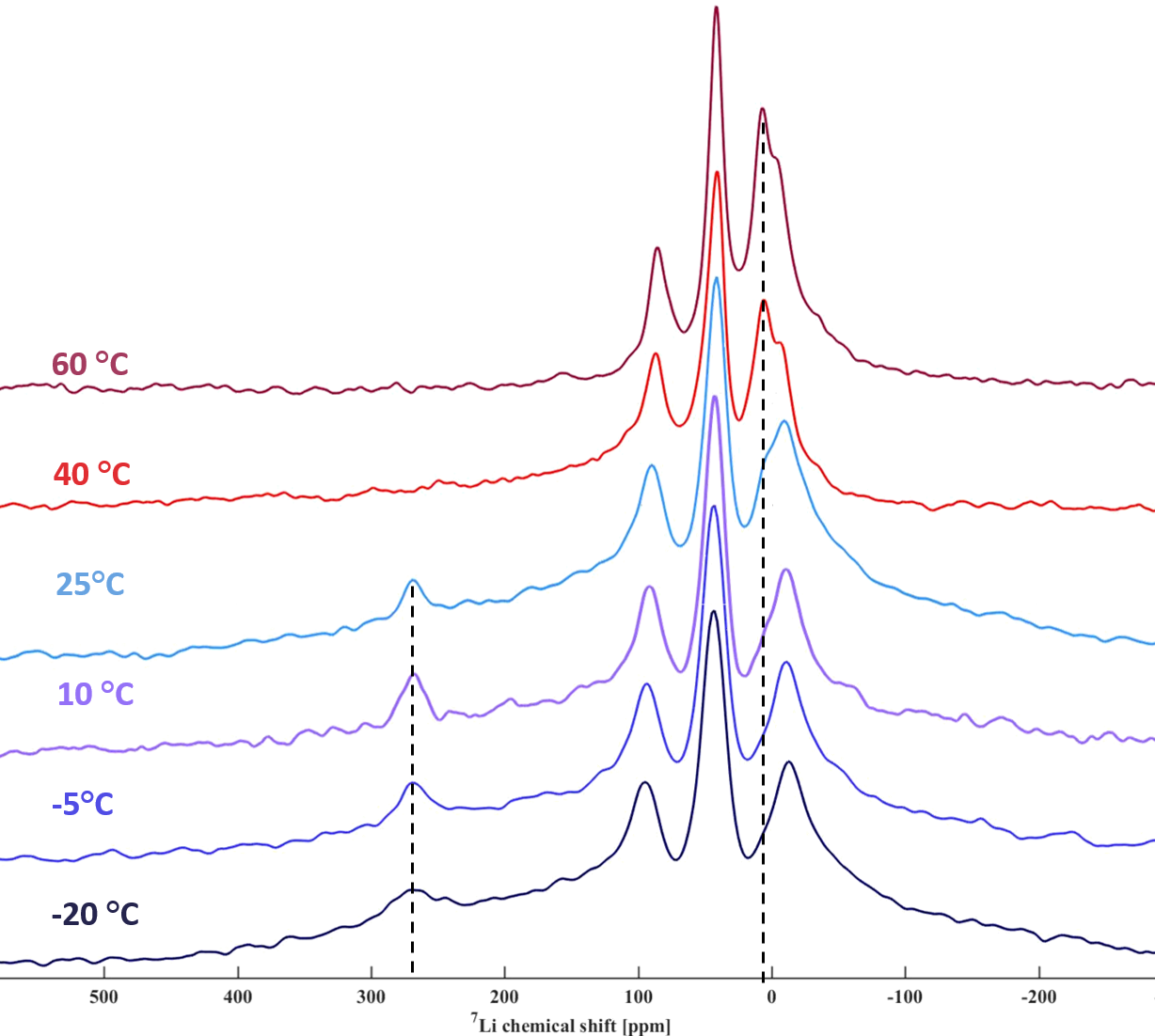}
\caption{\textbf{Representation of the {\ce{^7}Li} static NMR in function of different temperatures during the heat treatment cycle of the sample after 5 months ageing.} \textsuperscript{7}Li-NMR spectra recorded from $-20\;^\circ$C to $60\;^\circ$C, show the evolution of the high-ppm signal and the appearance of the low-ppm feature at {\em ca.} 10-13 ppm, marked both by dashed lines.}
\label{allT}
\end{figure}

Temperature dependent static \ce{^7Li} NMR of the lithiated HOPG sample was performed after 5 months of ageing of the cleaned sample, after removal of all surface decomposition products. The spectra were recorded with increasing temperature from $-20\;^\circ$C to $60\;^\circ$C in steps of $15\;^\circ$C. For temperature adjustment, the integrated system for the temperature control of the Bruker DIFFBB probe was used. Fig.~\ref{allT} shows the temperature series of the NMR spectra during heating of the sample.
Interestingly, the residual high-ppm peak, discussed in the main text, disappears, while concomitantly a peak at {\em ca.} 10--13~ppm rises. In addition, attention has to be paid to a broad background signal of the spectra between 130--300~ppm at temperatures from $-20\;^\circ$C to $25\;^\circ$C.
This background signal disappears at $40-60\;^\circ$C. Upon cooling to $25\;^\circ$C, it reappears and, correlated to this intensity change, the peak at 10--13 ppm decreases in intensity. This indicates a reversible character of the associated species. The high-ppm feature, however shows an irreversible character ({\em cf.} main text). 
The background entirely vanishes after an additional ageing of two months ({\em cf.} Fig.~6 in the main text). After the full seven months of ageing, the sharp peak at {\em ca.} 11~ppm exhibits a quadrupolar pattern of {\em ca.} 40~kHz. This and the overall narrower linewidths indicate that the sample reached a more ordered equilibrium state, as visible in Fig.~6 (main text, violet curve).

\section{SI 2. {\em Ab initio} thermodynamics}

\subsection{General concept}
The idea behind {\em ab initio} atomistic thermodynamics is to evaluate relative stabilities of different stoichiometries as a function of the chemical potential of the variable components (in our case, $\mu_{\mathrm{Li}}$), by expressing them in terms of quantities directly accessible via first principle calculations. Our system is effectively graphite in equilibrium with a reservoir of liquid Li, so we can in principle express free energies and free energy variations as a function of $\mu_{\mathrm{Li}}$. Following the same well established reasoning for surfaces,\cite{Rogal2006} we can partition the total free energy of the entire system as 
\begin{equation}
 G = G_{\mathrm{graphite}} + G_{\mathrm{lithium}} + \Delta G^{\mathrm{interc}}
\end{equation}

where graphite and lithium are extended (infinite) regions and $\Delta G_{\mathrm{interc}}$ accounts for the (finite) intercalation region. Then we may write, for any lithiated compound in the intercalation region, with $N_{\mathrm{Li}}$ Li atoms and $N_{\mathrm{C}}$ C atoms per \ce{C_6} formula:

\begin{align}
\Delta G^{\mathrm{interc}} &= G - G_{\mathrm{graphite}} - G_{\mathrm{lithium}} = \\
&= G(T, P, N_{\mathrm{C}}, N_{\mathrm{Li}}) - N_{\mathrm{C}}g_{\mathrm{C}}(T, P) - N_{\mathrm{Li}}\mu_{\mathrm{Li}}(T, P)\;,
\end{align}

where $g$ denotes the partial molar free energy, or, for an infinite reservoir such as liquid Li, the chemical potential. If we now introduce the limit case

\begin{equation}
\Delta G^{\mathrm{interc}}_{\mathrm{empty}} = G(T, P, N_{\mathrm{C}}, 0) - N_{\mathrm{C}}g_{\mathrm{C}}(T, P)
\end{equation}

as the ``formation energy'' of empty graphite in the intercalation region (in other words, a reference finite slab of empty graphite), then we can express

\begin{align}
\Delta\Delta G^{\mathrm{interc}} &= \Delta G^{\mathrm{interc}} - \Delta G^{\mathrm{interc}}_{\mathrm{empty}} = \nonumber \\
&= G(T, P, N_{\mathrm{C}}, N_{\mathrm{Li}}) - G(T, P, N_{\mathrm{C}}, 0) - N_{\mathrm{Li}}\mu_{\mathrm{Li}}(T, P)\;.
\label{deltadelta}
\end{align}

Trivially, this reference will be zero at any T, P and will allow us to conveniently evaluate the relative stability of any lithiated compound simply with respect to empty graphite. In that regard, we simply indicate those as $\Delta G$ rather than $\Delta \Delta G$ in the main text.

\subsection{Lithium chemical potential}
In general, we can express the chemical potential of lithium at any ($T$, $P$) as

\begin{equation}
 \mu_{\mathrm{Li}} (T, P) = \mu_{\mathrm{Li}}(0, 0) + \Delta \mu_{\mathrm{Li}}(T, P)
\end{equation}

and use the fact that at vanishing temperature and pressure, the chemical potential is equal to the internal energy $U$ of the stable state at $T = 0$~K ({\em i.e.} BCC lithium):

\begin{equation}
 \mu_{\mathrm{Li}}(0, 0) = g_{\mathrm{Li}} (0, 0) = U_{\mathrm{Li}} (0, 0) \simeq E^{\mathrm{DFT}}_{\mathrm{Li, BCC}}\;,
\end{equation}

where we approximate the internal energy as the calculated DFT total energy (neglecting the zero point energy). The variation $\Delta \mu_{\mathrm{Li}}(T, P)$ may be evaluated from {\em ab initio} calculations, or, where available, extracted from thermochemical tables\cite{janaf} as:

\begin{align*}
 \Delta \mu_{\mathrm{Li}}(T, P) &= \Delta G_{\mathrm{Li}} (T, P) = \Delta H - T \Delta S = \\
 &= \left[ H (T, P) - H (0, 0) \right]- T \left[S(T, P) - S (0, 0)\right] \;.
\end{align*}

Technically, thermochemical tables only report values at the standard pressure of $ P^0 = 0.1$ MPa rather than the hypothetical $P = 0$. 
However, as will be apparent in the next section, up to $P^0$ the effect of pressure on the chemical potential of condensed phases is negligible, so we can safely approximate $H (0, 0) = H (0, P^0)$ and $S(0,0) = S(0, P^0)$, obtaining:

\begin{equation}
 \Delta \mu_{\mathrm{Li}}(T, P^0) \simeq \left[ H (T, P^0) - H (0, P^0)\right] - T \left[S(T, P^0) - S (0, P^0)\right]\;.
 \label{noP}
\end{equation}

In principle, we may use the above equation to directly obtain the value of $\Delta \mu_{\mathrm{Li}}(T, P^0)$ at the experimental temperature $T = 500$ K. However, this would require extrapolating $H$ and $S$ to $T = 0$ K for {\em liquid} lithium, as those are not directly included in the corresponding thermochemical table -- contrary to the reference solid state. Thus, it is safer to calculate $\Delta \mu_{\mathrm{Li}}$ for solid lithium up to the fusion temperature $T_{\mathrm{fus}} = 453.69$ K and then, using the fact that, at the fusion temperature:

\begin{equation}
 \mu_{\mathrm{Li, sol}} = \mu_{\mathrm{Li, liq}} \;,
\end{equation}

proceed to calculate the chemical potential of liquid lithium for higher temperatures from there. Since the experimental temperature is not much larger than the fusion temperature, we can from this point on use a simple linear approximation for the variation of the chemical potential:

\begin{equation}
\Delta (T, P) = \mu(T_{\mathrm{fus}}, P^0) - s \Delta T + v \Delta P
\end{equation}

with $s$ is the negative molar entropy (also tabulated) and $v$ is the molar volume, which can be easily calculated from the experimental density.

Of note, it is easy to verify that $s \Delta P$, for $\Delta P = 0.1$ MPa, only changes the chemical potential by $10^{-5}$ eV, justifying the approximation in Eqn.~\ref{noP}. Conversely, at the much higher pressures usually employed in the synthesis of \ce{LiC2}, the effect is in the order of 1 eV. We note in passing that this straightforwardly brings the formation free energy \ce{LiC2} in the thermally accessible range at high pressure, in agreement to its reported synthesis.

\subsection{Free energies of graphite and \ce{Li_xC_y}}

At this point we have a full, quantitative expression for $\mu_{\mathrm{Li}}$ at any finite T and P. Let us now go back to Eqn.~\ref{deltadelta} and let us simplify the notation posing $N_{\mathrm{Li}} = x$ and $N_{\mathrm{C}} = y$ for a generic \ce{Li_xC_y} intercalation compound. As a first approximation (often used in AITD), one may neglect all $T$ and $P$ effects on the intercalated compounds, thus replacing the free energy with the DFT total energy: 

\begin{equation}
G_{\mathrm{Li_xC_y}}(T, P, x, y) \simeq U_{\mathrm{Li_xC_y}} (0, 0) \simeq E^{\mathrm{DFT}}_{\mathrm{Li_xC_y}}\;,
\end{equation}

and analogously for empty graphite. This approximation is extremely convenient and has been employed successfully for (semi-)quantitative predictions especially in heterogeneous catalysis, where it is largely justified by the fact that {\em i)} there the chemical potential which is allowed to vary is typically of some gas species, hence it is much more sensitive to $T$ and $P$ effect than condensed phases, and {\em ii)} it is assumed that the competing species are similar enough that the vibrational, entropic and PV contributions between different species largely cancel each other.\cite{Rogal2006} This allows to draw simple diagrams in which the relative stabilities of different stoichiometries vary linearly with the chemical potential of the variable component, with slopes proportional to its content in the formula. However, the neglection of the finite temperature contributions to the total energies should be carefully evaluated case by case -- as it turns out, they do make a difference in our case.

In the most complete picture, and employing the usual approximation of expressing the zero-temperature internal energy as the DFT total energy, one may write

\begin{equation}
 G = F + PV = \simeq E^{\mathrm{DFT}} + F_{vib} + PV;
\end{equation}

where $F_{vib}$ is the vibrational free energy, directly calculated from the first-principles vibrational density of states in the harmonic approximation.\cite{phonopy} The latter includes the Zero Point Energy, the vibrational contribution to the internal energy and the vibrational entropy. This, the only contribution neglected here is the configurational entropy. Its estimation is computationally intensive. For the scope of this work we may limit ourselves to qualitatively note that, being by definition always positive or zero, it would be positive for lithiated compounds and zero for pristine graphite and perfect \ce{LiC2}, for which only one realisation is possible. Therefore, it would lower the formation free energy for all the intermediate compounds. 

All the calculated free energies are normalized per graphite unit (or, equivalently and consistently with the notation used here, per mole of \ce{Li_xC_y} with $y=6$).

\section{SI 3. {\em Ab initio} Molecular Dynamics}

\subsection{Diffusion coefficients and effective diffusion barrier}

Due to the high computational cost of DFT, MD was performed based on Density Functional Tight Binding (DFTB), a semi-empirical tight-binding approximation to DFT, with the recently developed parametrisation\cite{Panosetti2021} augmented with Li--Li repulsion.

The diffusion coefficients were calculated from the mean square displacements using Einstein's relation

\begin{equation}
 D = \dfrac{1}{2nt} \left \langle \left | r(t) - r(0) \right | ^{2} \right \rangle
\end{equation}

where $r(t)$ is the position at time $t$ and $n$ is the number of degrees of freedom. Only the 2D diffusion in the $xy$ plane was taken into account. The analysis was performed using a module implemented in ASE, using an average over 3 segments per trajectory to improve statistics.

\begin{figure}[h]
\includegraphics[width=.95\linewidth]{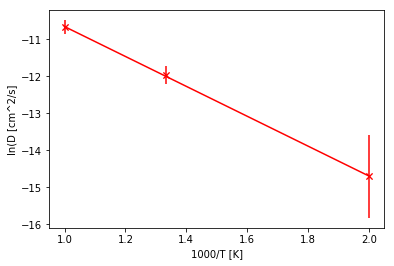}
\caption{Arrhenius plot of diffusion coefficients at 500, 750 and 1000~K. The error bars are computed by error propagation based on the standard deviation of the mean square displacement. The error for 500~K is larger due to the limited statistics, due to the fact that at lower temperatures the trajectory is dominated by vibrational noise with very few true diffusion events (jumps).}
\label{arrhenius}
\end{figure}

The diffusion coefficients are reported in table\ref{diff} and plotted in Fig.~\ref{arrhenius}.

\begin{table}[h]
\caption{Diffusion coefficients from MD}
\label{diff}
\begin{tabular}{cccc}
 \hline
 $T$ ($K$) & 500 & 750 & 1000 \\
 $D \pm \sigma$ ($cm^2/s$) & $4.1\cdot 10^{-7} \pm 4.6 \cdot 10^{-7}$ & $6.3 \cdot 10^{-6} \pm 1.5 \cdot 10^{-6}$ & $2.3 \cdot 10^{-5} \pm 4.4 \cdot 10^{-6}$ \\ 
 \hline
\end{tabular}
\end{table}

The ratio between characteristic times $t_2$ and $t_1$ at temperatures $T_2$ and $T_1$ can be calculated as

\begin{equation}
 \dfrac{t_2}{t_1} = \dfrac{k_1}{k_2} =\exp\left[\dfrac{(T_1-T_2)\cdot E_a}{k_B\cdot T_1 \cdot T_2}\right] 
\end{equation}

where $k_1$ and $k_2$ are the rates at $T_1$ and $T_2$ respectively, $E_a$ is the effective activation barrier of the diffusion, and $k_B$ is the Boltzmann constant.

\subsection{Degradation of clusters from MD snapshots}

\begin{figure}[h]
\includegraphics[width=.9\linewidth]{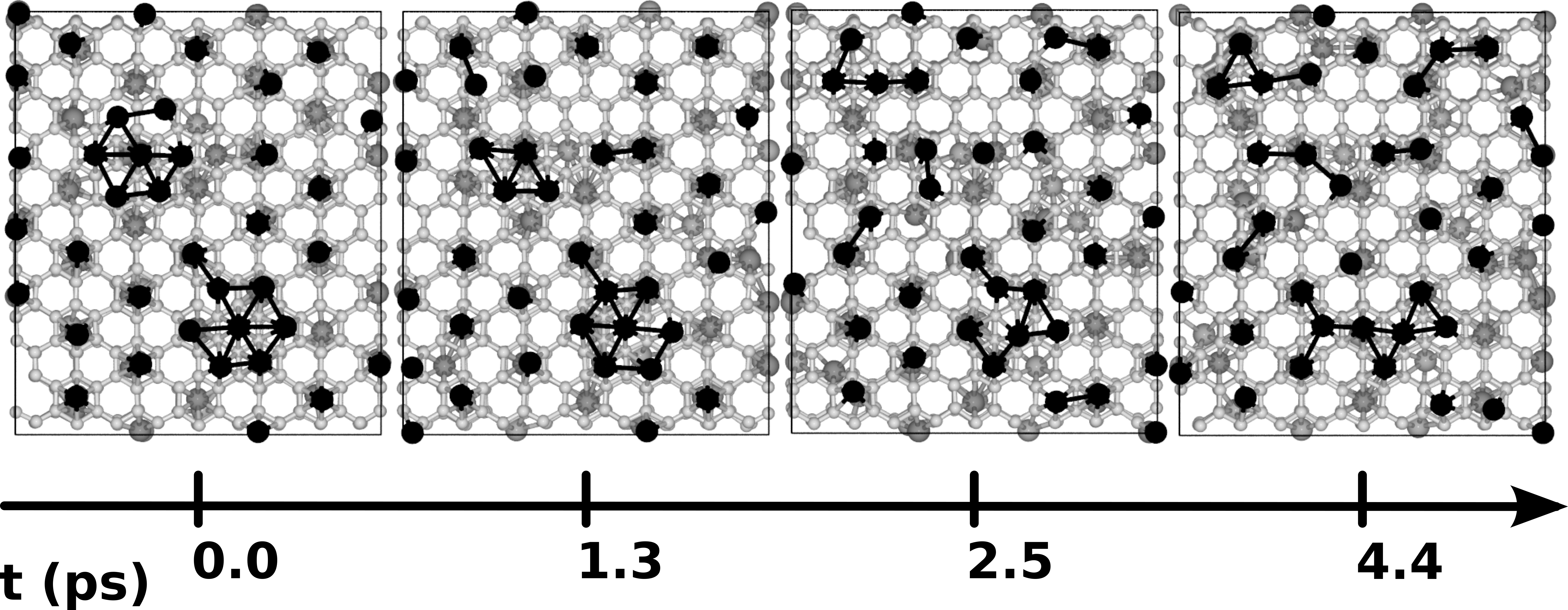}
\caption{Selected snapshots in the MD trajectory at 750~K showing the frequent appearance of Li-Li dimers as transient byproducts of superdense cluster decomposition.}
\label{md_snaps_750}
\end{figure}

Fig.~\ref{md_snaps_750} shows selected snapshots of the MD trajectory, at 750~K to show advanced stages of degradation of \ce{Li7} clusters. Similar fragments appear in the 500~K trajectory, only significantly more slowly. The occurrence of disordered planar clusters with various coordinations, as well as dimers and trimers is evident.

\bibliography{bibliography.bib}